\documentclass[twocolumn, 10pt]{IEEEtran}
\usepackage{url}
\usepackage[breaklinks]{hyperref} % create hyperlinks
\usepackage{breakurl}
\usepackage{graphicx}
\usepackage{amssymb}
\usepackage{amsmath}
\usepackage{mathtools}
\usepackage{cite}
\usepackage{stfloats}
\usepackage{epstopdf}
\usepackage{psfrag}
\usepackage[mathscr]{euscript}
\usepackage{acronym}  % make an acronym
\usepackage{booktabs}
\usepackage{mdframed}
\usepackage[table]{xcolor}
\usepackage{multirow}
\usepackage{rotating}
\usepackage{verbatim}
\usepackage[skip=0.333\baselineskip]{caption}
\usepackage{dcolumn}
\usepackage{amsfonts}
\usepackage{makecell}
\usepackage{array}
\usepackage{hhline}
\usepackage{ctable}
\usepackage{caption}
\usepackage{subcaption}

\newcommand{\block}{S_{\text{b}}}
\newcommand{\timeout}{T_{\text{b}}}
\newcommand{\TVarAvg}{\overline{T}_{\text{t}}}
\newcommand{\txnrate}{\lambda_{\text{t}}}
\newcommand{\TVarep}{\overline{T}_{\text{ep}}}
\newcommand{\TVarbf}{\overline{T}_{\text{bf}}}

\newcommand{\Tt}{T_{\text{t}}}
\newcommand{\Te}{T_{\text{e}}}
\newcommand{\To}{T_{\text{o}}}
\newcommand{\Tv}{T_{\text{v}}}
\newcommand{\Deltat}{\Delta_t}
\newcommand{\expectation}{\mathbb{E}}
\newcommand{\alphao}{\alpha_{\text{o}}}
\newcommand{\betao}{\beta_{\text{o}}}
\newcommand{\alphat}{\alpha_{\text{t}}}
\newcommand{\betat}{\beta_{\text{t}}}
\newcommand{\Ti}{T_{i}}
\newcommand{\alphai}{\alpha_{i}}
\newcommand{\betai}{\beta_{i}}

\acrodef{HLF}{Hyperledger Fabric}
\acrodef{PDF}{probability density function}
\acrodef{CDF}{cumulative distribution function}
\acrodef{PBFT}{practical Byzantine fault tolerance}
\acrodef{SRN}{stochastic reward net}
\acrodef{AP}{access point}
\acrodef{BS}{base station}
\acrodef{KS}{Kolmogorov-Smirnov}
\acrodef{BFT}{Byzantine fault tolerance}
\acrodef{IoT}{Internet of Things}
\acrodef{UAV}{unmanned aerial vehicle}
\acrodef{SDN}{software-defined networks}
\acrodef{IIoT}{industrial IoT}
\acrodef{HeApp}{HLF-enabled application}
\acrodef{GEV}{generalized extreme value}
\acrodef{GSPN}{generalized stochastic petri net}

\begin{document}
	
\newcommand{\paperTitle}{Facing to Latency of Hyperledger Fabric for \\ Blockchain-enabled IoT: Modeling and Analysis}
	
%---------------------------------------------------------------------------%
%                     title, title footnote, header                         %
%---------------------------------------------------------------------------%

%\twocolumn

% paper title
\title{\paperTitle}
% \title{Distributed Secrecy in \\Multilevel Wireless Networks}

% author names, IEEE memberships, corresponding address, title footnote %
\author{
	%%%%%%%% [begin] %%%%%%%%
	\vspace{0.2cm}
	Sungho Lee, Minsu Kim, Jemin Lee, \textit{Member, IEEE}, Ruei-Hau Hsu, \textit{Member, IEEE}, 
	\\
	Min-Soo Kim, and Tony Q. S. Quek, \textit{Fellow, IEEE}
	
	\thanks{
		Corresponding author is J. Lee.
		
		S.\ Lee is with the Department of Information and Communication Engineering, Daegu Gyeongbuk Institute of Science and Technology (DGIST), Daegu 42988, Republic of Korea (e-mail: \texttt{seuho2003@dgist.ac.kr}).
		
		M. Kim is with the Wireless@VT Group, Bradley Department of Electrical and Computer Engineering, Virginia Tech, Blacksburg, VA 24061 USA (e-mail: \texttt{msukim@vt.edu})
		
		J.\ Lee is with the Department of Electrical and Computer Engineering, Sungkyunkwan University (SKKU), Suwon 16419, Republic of Korea (e-mail: \texttt{jemin.lee@skku.edu}).
		
		R.\ -H.\ Hsu is with the Department of Computer Science and Engineering, National Sun Yat-sen University, Kaohsiung 80424, Taiwan, R.O.C.
		(e-mail: \texttt{rhhsu@mail.cse.nsysu.edu.tw}).
		
		M.\ -S.\ Kim is with the School of Computing, Korea Advanced Institute of Science and Technology (KAIST), Daejeon 34141, Republic of Korea
		(e-mail: \texttt{minsoo.k@kaist.ac.kr}).
		
		T.\ Q.\ S.\ Quek is with Information Systems Technology and Design Pillar, 
		Singapore University of Technology and Design, Singapore 487372 
		(e-mail: \texttt{tonyquek@sutd.edu.sg}).
	}
	%[-0.5em]
%
%
%
}
%% make the title area
%% Don't write page number 0 to the cover page.
\maketitle %% make the title area
%% Don't write page number 0 to the cover page.

%
% \markboth{Submitted to IEEE Journal on Selected Areas in ommunications}{\title}

%
%%%%%%%%% uncomment this section for a 2-column formt %%%%%%%
%%%%%%%%% [begin] %%%%%%%%
%\thispagestyle{empty}
%  \textcolor{blue}{\framebox{\textsf{\small{Today: \today}}}}\\

%
%\newpage
%%%%%%%%% [end] %%%%%%%%
\setcounter{page}{1}
\acresetall
%%---------------------------------------------------------------------------%
%%                           abstract and key words                          %
%%---------------------------------------------------------------------------%

\begin{abstract}
Hyperledger Fabric (HLF), one of the most popular private blockchains, has recently received attention for blockchain-enabled Internet of Things (IoT). However, for IoT applications to handle time-sensitive data, the processing latency in HLF has emerged as a new challenge. In this article, therefore, we establish a practical HLF latency model for HLF-enabled IoT. We first discuss the structure and transaction flow of HLF-enabled IoT. After implementing real HLF, we capture the latencies that each transaction experiences and show that the total latency of HLF can be modeled as a Gamma distribution, which is validated by conducting a goodness-of-fit test (i.e., the Kolmogorov-Smirnov (KS) test). We also provide the parameter values of the modeled latency distribution for various HLF environments. Furthermore, we explore the impacts of three important HLF parameters including transaction generation rate, block size, and block-generation timeout on the HLF latency. As a result, this article provides design insights on minimizing the latency for HLF-enabled IoT.
\end{abstract}

%\begin{IEEEkeywords}
%	Blockchain, Hyperledger Fabric, industrial Internet of Things, probability distribution fitting, latency
%\end{IEEEkeywords}

\section{Introduction}\label{sec:Introduction}
Since blockchain technology was developed as a novel solution to ensure data integrity in distributed systems, various blockchain-enabled \ac{IoT} applications have recently emerged. In the blockchain-enabled \ac{IoT} applications, delivered \ac{IoT} data is appended to the assigned remote blockchain storage. Blockchains can be classified into two types: public blockchain and private blockchain. The private blockchain has recently received much attention in terms of fast transaction processing and privacy preservation \cite{MJo:20}. Particularly, \ac{HLF}\footnote{\url{https://www.hyperledger.org/use/fabric}.}, which is hosted by Linux Foundation and contributed by IBM since 2015, is gaining popularity \cite{DaiZheZha:19} as one of the promising private blockchains. As discovered in literature \cite{DixSinRahRaj:21, JamAhmIqbKim:20, AggKumAlhMuh:21}, \ac{HLF}-enabled \ac{IoT} applications include \ac{IIoT}, healthcare, wireless monitoring, and \acp{UAV}. For example, decentralized \ac{IoT} data marketplace is proposed using \ac{HLF} in \cite{DixSinRahRaj:21}. For smart healthcare, \ac{HLF} is exploited for \ac{IoT} to share monitored patient vital signs \cite{JamAhmIqbKim:20} and to ensure security and privacy for medical data in the Healthcare 4.0 industry using \acp{UAV} \cite{AggKumAlhMuh:21}. Thus, it is expected that \ac{HLF} will be more widely used to securely manage a high volume of \ac{IoT} data.

However, in blockchain platforms including \ac{HLF}, high latency is still recognized as a major issue \cite{DaiZheZha:19}. Specifically, certain \ac{IoT} applications, which are handling time-critical data, may have strict latency requirements for making correct and useful decision \cite{WeiJorSahNik:14}, and require fast transaction commitment. For example, vital signs of patients need to be promptly processed and transferred to smart hospital blockchain networks for effective medical care \cite{JamAhmIqbKim:20}. Therefore, for those applications, it is important to predict and understand the \ac{HLF} latency, prior to their practical implementations. Nonetheless, in most works on \ac{HLF}-enabled applications, the \ac{HLF} latency is generally not considered or simply ignored, and the latency characteristics including its distribution have not been fully explored.

Recently, there have been some works that analyze the \ac{HLF} latency such as the ones for \ac{HLF} v0.6 \cite{SukMarChaTriRin:17} and the ones for \ac{HLF} v1.0 or higher \cite{SukWanTriRin:18, YuaZheXioZhaLei:20, JiaChaLiuMisMis:20, XXu:21}.\footnote{Note that \ac{HLF} v0.6 has a different structure from that of \ac{HLF} v1.0 or higher.} Specifically, a \ac{SRN} model is used to define the latency of \ac{HLF} v0.6 in \cite{SukMarChaTriRin:17}. For \ac{HLF} v1.0 or higher, by considering Kafka/ZooKeeper, which is a new consensus algorithm released in \ac{HLF} v1.0, the latency is defined and measured based on several theoretical models including \ac{SRN} \cite{SukWanTriRin:18}, \acp{GSPN} \cite{YuaZheXioZhaLei:20}, a hierarchical model based on transaction execution and validation \cite{JiaChaLiuMisMis:20}, and a queueing-based model \cite{XXu:21}.

However, in all above \ac{HLF} latency studies, only the mean latency is presented without showing the latency distribution, and in most works, the theoretical models are not verified in real \ac{HLF}. When the latency is measured from real implementation of \ac{HLF}, it is quite random, so the distribution of the latency is importantly required for the design of reliable \ac{HLF}-enabled \ac{IoT}. Furthermore, \ac{HLF} has a sophisticated structure, where many network parameters affect the latency simultaneously, so their impacts on the latency should be carefully investigated to lower the \ac{HLF} latency.

In this article, therefore, we establish a latency model of \ac{HLF} especially for \ac{HLF}-enabled \ac{IoT}, where time-sensitive data is generally managed. After characterizing the latency distribution, we also analyze the impacts of three important \ac{HLF} parameters (i.e., block size, block-generation timeout, and transaction generation rate) on the \ac{HLF} latency. We then discuss how the \ac{HLF} latency can be reduced by setting those parameters properly. The contributions of this article can be summarized as follows:

\begin{itemize}
	\item We newly provide the probability distribution of the \ac{HLF} latency by performing the probability distribution fitting on transaction latency samples, which are captured in real \ac{HLF}. We also provide the parameter values of corresponding probability distributions for various \ac{HLF} parameter setups. To the best of our knowledge, this is the first work that provides the distribution of the \ac{HLF} latency.

	\item We develop the latency model of \ac{HLF} with a new structure, which is adopted in the latest \ac{HLF} releases. Hence, our work is more compatible with current \ac{HLF}-enabled \ac{IoT}, compared to the previous \ac{HLF} latency model for \ac{HLF} v0.6 \cite{SukMarChaTriRin:17}.
	
	\item We explore the impacts of three main \ac{HLF} parameters (i.e., block size, block-generation timeout, and transaction generation rate) on the \ac{HLF} latency. Especially, we figure out the existence of the best values of those \ac{HLF} parameters that minimize the \ac{HLF} latency. This provides some design insights on \ac{HLF} parameters to lower the \ac{HLF} latency for more reliable \ac{HLF}-enabled \ac{IoT}.
\end{itemize}

%The remainder of this article is organized as follows. In Section \ref{sec:HLFOverview}, we describe a structure of \ac{HLF}-enabled \ac{IoT} and outline the transaction flow in \ac{HLF}. We then provide our \ac{HLF} latency model based on probability distribution fitting in Section \ref{sec:HLFLatencyModeling}. In Section \ref{sec:ParameterAnalysis}, we analyze the impacts of the important parameters aforementioned on the \ac{HLF} latency, and then explore the methods for minimizing the mean \ac{HLF} latency. Lastly, the conclusions of this article are summarized in Section \ref{sec:Conclusions}.

\section{HLF-enabled IoT}\label{sec:HLFOverview}
\ac{HLF} is a private blockchain platform for a modular architecture, which is one of the sub-projects in the Hyperledger Project. In the other blockchain platforms, a deterministic programming language is compulsory to prohibit their ledgers from diverging, such as Solidity in Ethereum. In contrast, the novel structure of \ac{HLF} v1.0 or higher not only enables general-purpose programming languages, but also prevents ledger bifurcation at the same time. In this section, we first describe a basic structure of \ac{HLF}-enabled \ac{IoT}. We then demonstrate the transaction flow consisting of three phases, and provide the definitions of some important parameters in \ac{HLF}. Note that more detailed explanations of \ac{HLF} are provided in \cite{EAnd:18}.

\subsection{\ac{HLF}-enabled \ac{IoT} Structure}
\begin{figure}[t!]
	\centering
	%\captionsetup{justification=centering}
	\begin{center}   
		{
			\includegraphics[width=1.00\columnwidth]{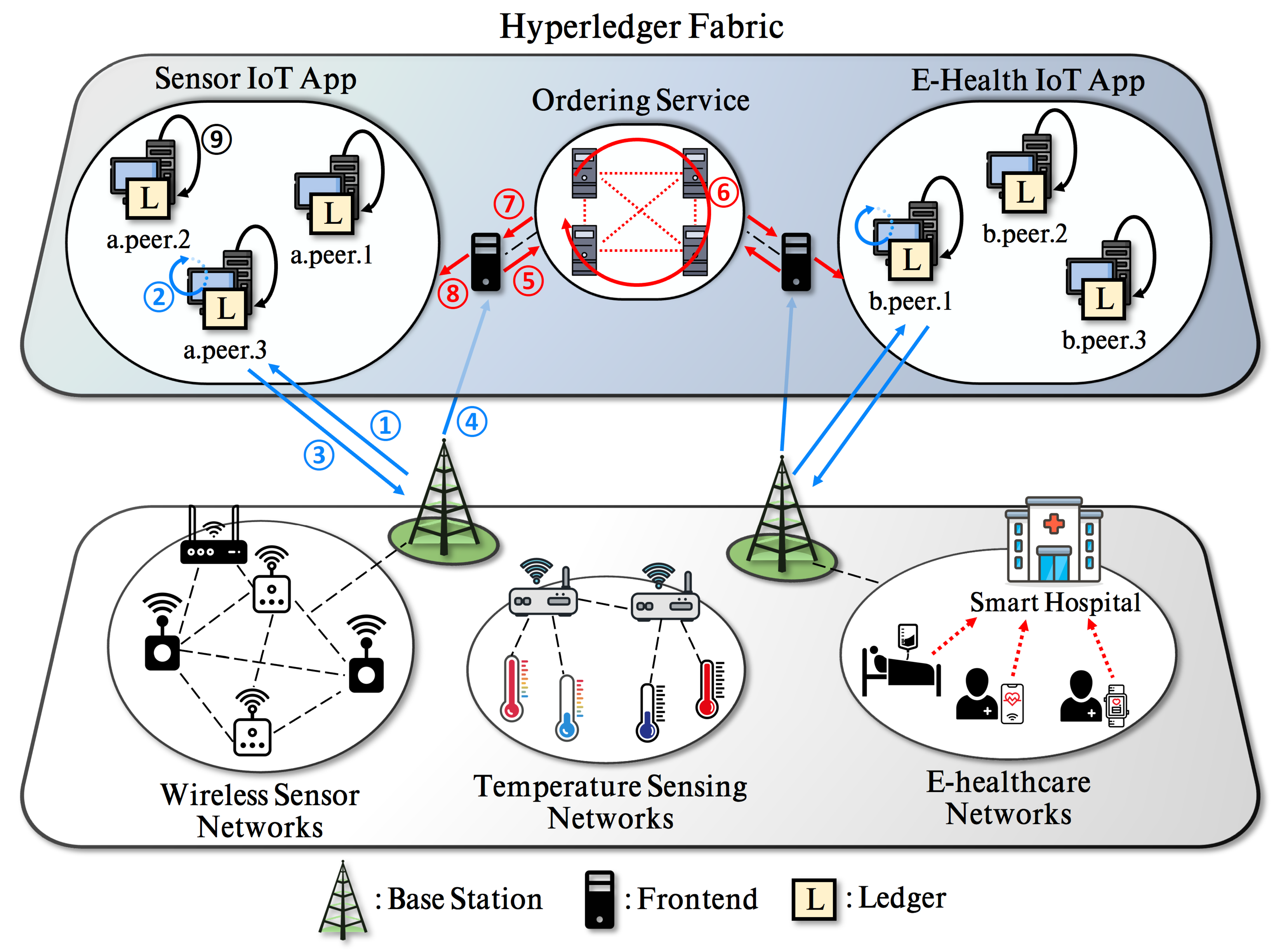}
			%			\vspace{-10mm}
		}
	\end{center}
	\caption{
		Structure of \ac{HLF}-enabled \ac{IoT}.
	}
	\vspace{-3mm}
	\label{fig:mainFig}
\end{figure}

Figure \ref{fig:mainFig} shows the basic structure of \ac{HLF}-enabled \ac{IoT}, where the data, delivered from \ac{IoT} devices, are stored and managed by \ac{HLF}. Note that \ac{HLF} can support multiple \ac{IoT} applications, but it can provide an independent ledger for each application for data confidentiality. Thus, each \ac{IoT} application can maintain its ledger without unauthorized access. As shown in Fig. \ref{fig:mainFig}, the data from an \ac{IoT} device is transmitted through a \ac{BS} or an \ac{AP}. The data is then appended to the ledger in \ac{HLF} by following the transaction flow, which will be described in Section \ref{subsec:TransactionFlow}.

\subsection{Transaction Flow}\label{subsec:TransactionFlow}
Transactions in \ac{HLF} are appended to the ledger through three phases: endorsing phase, ordering phase, and validation phase. The processing phases enable \ac{HLF} networks not only to support general-purpose programming languages, but also to remain tamper-proof. In this subsection, we describe the transaction flow in \ac{HLF} for data updates.

\subsubsection{Endorsing Phase}
The endorsing phase is to ensure that endorsing peers have the same transaction execution results obtained from each of their copied ledgers. This phase is necessary to prevent blockchain forks. If they are identical, the transaction is delivered to the ordering service. In contrast, the transaction is discarded if they are different.

\subsubsection{Ordering Phase}
The ordering phase is to order all transactions chronologically and generate new blocks in the ordering service. Note that the ordering service is a node cluster, where a pre-defined consensus protocol is conducted. A newly generated block is transmitted to peers for the validation phase through the gossip protocol. Note that we focus on the Kafka/ZooKeeper-based ordering service, which is currently the most widely used \cite{LeeKimLeeHsuQue:21}.

\subsubsection{Validation Phase}
The validation phase is the last phase in which the new block is validated by each peer individually. In this phase, a receiving peer mainly conducts sequential verification processes on each transaction in the block, in the order described in \cite{EAnd:18}. If any of the verification processes is violated at a transaction, it is marked as invalid and prohibited from updating the ledger.

\subsection{Important HLF Parameters}\label{subsec:HLFParameters}
In this subsection, we define three important \ac{HLF} parameters, which mainly affect the \ac{HLF} latency.

\subsubsection{Block Size}
The block size $\block$ is one of the configurable parameters of the ordering service. This parameter defines the maximum number of transactions in one block. In other words, the block size determines how many transactions the ordering service can collect the most in one block. The transactions are promptly exported from the ordering service as a new block, when the number of transactions in the waiting queue reaches a pre-defined block size value.

\subsubsection{Block-generation Timeout}
The block-generation timeout $\timeout$ is another configurable parameter of the ordering service. This parameter defines the maximum time to wait for other transactions to be exported as a new block since the first transaction arrived at the waiting queue. Once the timer expires, the waiting transactions are included into a new block, regardless of their number in the waiting queue.

\subsubsection{Transaction Generation Rate}
The transaction generation rate $\txnrate$ refers to the number of transactions arrived in \ac{HLF} per second. Being different from the two configurable parameters above, this parameter can be determined by many factors. For example, when an \ac{IoT} device conveys data to \ac{HLF} as transactions, the communication channel and data generation frequency of the device can affect this parameter.

\begin{figure*}[t!]
	\centering
	\begin{subfigure}[b]{0.245\textwidth}
		\centering
		\captionsetup{justification=centering}
		\psfrag{XXX}[Bc][bc][0.8] {$t$ [sec]}
		\psfrag{YYY}[Bc][bc][0.8] {Probability/$\Deltat$}
		%\psfrag{Hist}[Bl][Bl][0.7] {Endorsing latency histogram}
		%\psfrag{DDDDDDDDDDDDDDDDDDDDDDDDDDDDD}[Bl][Bl][0.7] {Exponential distribution PDF [$\lambda=94.5$]}
		\includegraphics[width=\textwidth]{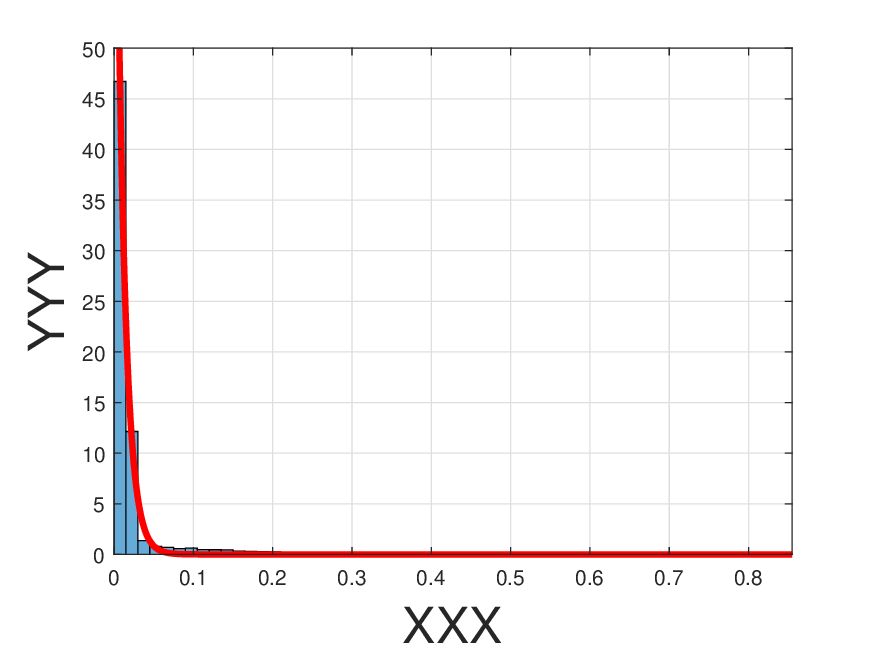}
		\caption{Endorsing latency:\\Exponential dist. PDF\\$[\lambda\hspace{-0.3mm}=\hspace{-0.3mm}94.5]$}
		\label{fig:el_t1b10r10}
	\end{subfigure}
	\hfill
	\begin{subfigure}[b]{0.245\textwidth}
		\centering
		\captionsetup{justification=centering}
		\psfrag{XXX}[Bc][bc][0.8] {$t$ [sec]}
		\psfrag{YYY}[Bc][bc][0.8] {Probability/$\Deltat$}
		%\psfrag{Hist}[Bl][Bl][0.7] {Ordering latency histogram}
		%\psfrag{DDDDDDDDDDDDDDDDDDDDDDDDDDDDDDDDDDD}[Bl][Bl][0.7] {Gamma distribution PDF [$\alpha=2.652$, $\beta=3.458$]}
		\includegraphics[width=\textwidth]{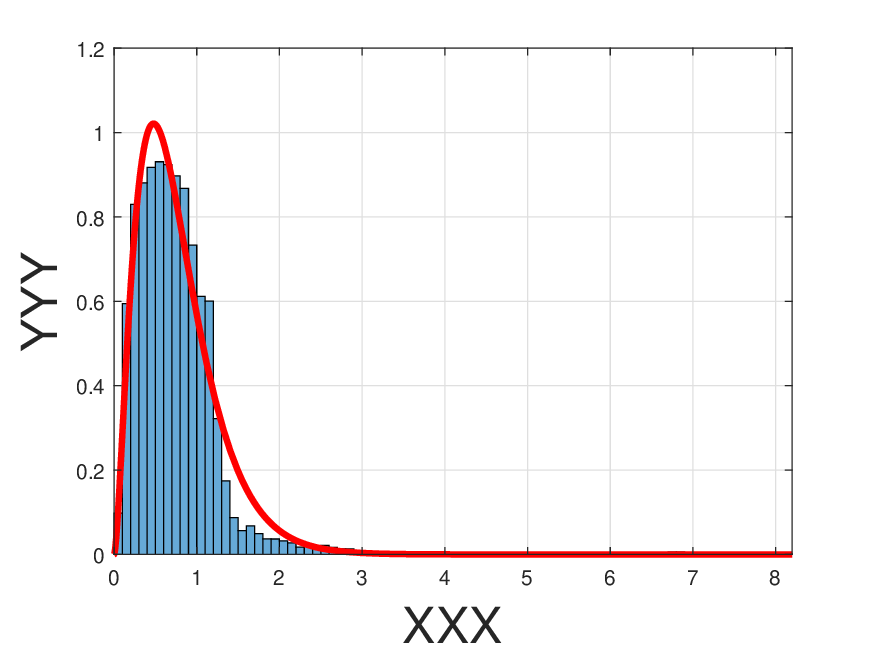}
		\caption{Ordering latency:\\Gamma dist. PDF\\$[\alphao\hspace{-0.3mm}=\hspace{-0.3mm}2.652, \betao\hspace{-0.3mm}=\hspace{-0.3mm}3.458]$}
		\label{fig:ol_t1b10r10}
	\end{subfigure}
	\hfill
	\begin{subfigure}[b]{0.245\textwidth}
		\centering
		\captionsetup{justification=centering}
		\psfrag{XXX}[Bc][bc][0.8] {$t$ [sec]}
		\psfrag{YYY}[Bc][bc][0.8] {Probability/$\Deltat$}
		\psfrag{Hist}[Bl][Bl][0.7] {Validation latency histogram}
		%\psfrag{D}[Bl][Bl][0.7] {GEV distribution PDF}
		%\psfrag{DDDDDDDDDDDDDDDDDDDDDDDDDDDl}[Bl][Bl][0.7] {[$\xi=0.2105$, $\sigma=0.1441$, $\mu=0.4797$]}
		\includegraphics[width=\textwidth]{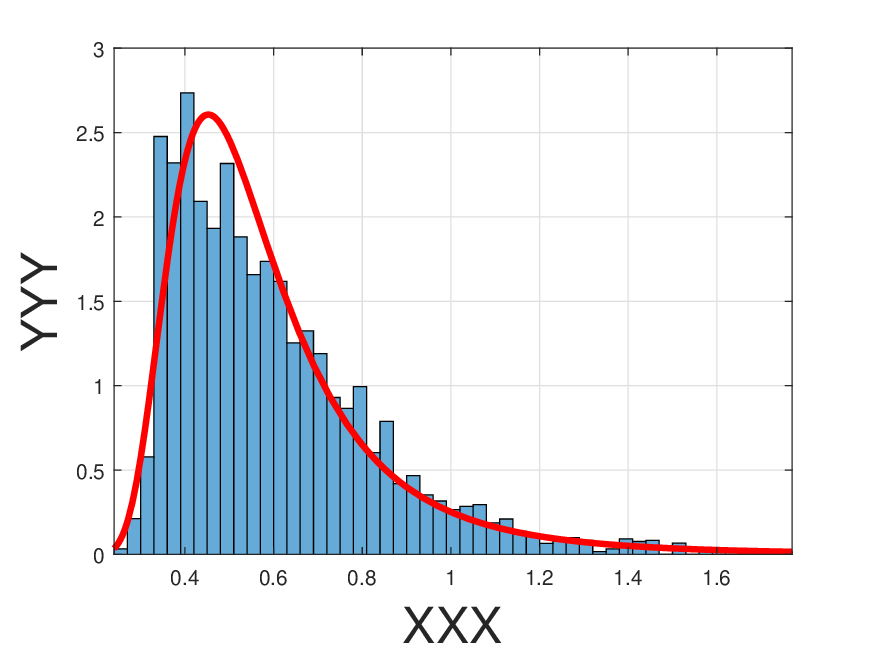}
		\caption{Validation latency:\\Fr\'{e}chet dist. PDF\\$[\alpha \hspace{-0.3mm} = \hspace{-0.3mm} 0.21, s \hspace{-0.3mm} = \hspace{-0.3mm} 0.479, m \hspace{-0.3mm} = \hspace{-0.3mm} 0.144]$}
		\label{fig:vl_t1b10r10}
	\end{subfigure}
	\hfill
	\begin{subfigure}[b]{0.245\textwidth}
		\centering
		\captionsetup{justification=centering}
		\psfrag{XXX}[Bc][bc][0.8] {$t$ [sec]}
		\psfrag{YYY}[Bc][bc][0.8] {Probability/$\Deltat$}
		%\psfrag{Hist}[Bl][Bl][0.7] {Ledger-commitment latency histogram}
		%\psfrag{DDDDDDDDDDDDDDDDDDDDDDDDDDDDDDDDDDDDDD}[Bl][Bl][0.7] {Gamma distribution PDF [$\alpha=7.362$, $\beta=5.445$]}
		\includegraphics[width=\textwidth]{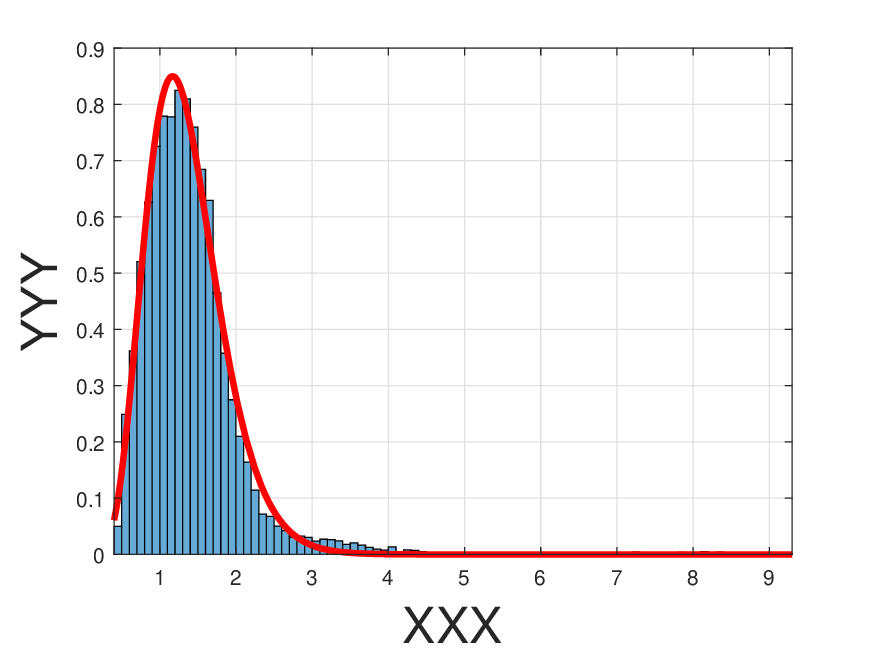}
		\caption{Total latency:\\Gamma dist. PDF\\$[\alphat\hspace{-0.3mm}=\hspace{-0.3mm}7.362, \betat\hspace{-0.3mm}=\hspace{-0.3mm}5.445]$}
		\label{fig:lc_t1b10r10}
	\end{subfigure}
	
	\caption{Histograms (blue bars) and their best-fit distributions (red lines) of the endorsing latency, ordering latency, validation latency, and ledger-commitment latency, where $\block=10$, $\timeout = 1$ [sec], and $\txnrate=10$ transactions/second.}
	\vspace{-3mm}
	\label{fig:lambda10}
\end{figure*}

\section{HLF Latency Modeling}\label{sec:HLFLatencyModeling}
In this section, we conduct \ac{HLF} latency modeling based on the probability distribution fitting method. We first define latency types in \ac{HLF}, and provide the experimental setup of our \ac{HLF} networks. For the probability distribution fitting method, we first capture transaction latency samples from the \ac{HLF} networks. We then the best fit distribution for each latency type, and finally provide our \ac{HLF} latency model. For evaluating goodness-of-fit of our modeling, we also conduct the \ac{KS} test. The feasibility conditions for our latency modeling are discussed with providing the cases, for which the probability distribution fitting method does not work well.

\subsection{Latency Types in \ac{HLF}}\label{subsec:HLFLatencyTypes}
Before modeling the \ac{HLF} latency, we first divide the total elapsed time of the data in the \ac{HLF} network according to the transaction-processing stages, described in Section \ref{subsec:TransactionFlow}, such as the endorsing latency, the ordering latency, the validation latency, and the ledger-commitment latency (i.e., the total latency). Note that we exclude the communication latency spent from an \ac{IoT} device to reach the \ac{HLF} network, as it is more depending on the communication standard that \ac{IoT} devices use, and it has already been explored in many literature.

\subsubsection{Endorsing Latency}
The endorsing latency refers to the time taken to request endorsing peers to endorse a transaction, and to receive their responses. To be specific, it is defined as the sum of the latencies for step 1, 2, 3, and 4 in Fig. \ref{fig:mainFig}.
	
\subsubsection{Ordering Latency}
The ordering latency refers to the time taken to await until the transaction is exported as a new block from the ordering service. It is defined as the sum of the latencies for step 5, 6, 7, and 8 in Fig. \ref{fig:mainFig}. Note that the latency for step 6 consists of the consensus time and the waiting time for other transactions to be released as part of a new block. In contrast, the latency for step 8 is the sum of transmission time and the waiting time of the released block for the validation phase. Thus, it may be considered as part of the validation latency. In this article, however, we include this latency into the ordering latency, when we depict the histograms in Figs. \ref{fig:lambda10} and \ref{fig:modelingCondition}. Note that the latency for step 8 may be highly likely to work as a bottleneck if the previous block requires long time to be validated. 
	
\subsubsection{Validation Latency}
The validation latency refers to the time taken to validate and commit the block with ledger update. This latency is defined as the latency for step 9 in Fig. \ref{fig:mainFig}. All transactions in the same block are sequentially validated, but committed to the ledger all together as a batch. Therefore, they have the identical validation latency from each other. Note that the validation latency may vary according to peer's type. In this article, this latency is measured at the endorsing peer.

\subsubsection{Ledger-commitment Latency (Total Latency)}
The ledger-commitment latency refers to the time taken to completely process a transaction from the beginning. Thus, it is the sum of the endorsing, ordering, and validation latencies.

\subsection{Experiment Setup}\label{subsec:ExperimentalSetup}
For the implementation of our \ac{HLF} network, we exploit the \ac{HLF} v1.3 sample network\footnote{\url{https://hyperledger-fabric.readthedocs.io/en/release-1.3}.} on one physical machine. The machine running the \ac{HLF} network is with Intel\textregistered~Xeon W-2155 @ 3.30 GHz processor and 16 GB RAM. In addition, we newly bring up a Kafka/ZooKeeper-based ordering service consisting of four Kafka nodes, three ZooKeeper nodes, and three frontend nodes that are connected to the Kafka node cluster in the network (i.e., the minimum number of required Kafka and ZooKeeper nodes for crash fault tolerance). Note that the frontend node is not only to inject transactions from clients into the Kafka node cluster, but also to receive new blocks, which will be disseminated to peers. The \ac{HLF} network has one endorsing peer and two committing peers. Note that the number of committing peers may not affect the ledger-commitment latency because they only validate newly delivered blocks and hold copied ledgers.

We perform 10 test runs for one experiment setup, and each test run includes 1,000 transaction proposals transmitted. All proposals attempt to update a key-value set, using appropriate parameters for the \textit{changeCarOwner} function defined in the \textit{Fabcar} chaincode\footnote{\url{https://github.com/hyperledger/fabric-samples/blob/release-1.3/chaincode/fabcar/go/fabcar.go}.}.
Each of the equipped frontend nodes has an equal probability of being selected to relay a transaction from our client to the Kafka node cluster, and to make a new block propagated. Note that the generated proposals in one test run do not access the same key-value set. Hence, we do not consider the multi-version concurrency control (MVCC) violation during the validation phase in this article.

New transactions are generated at the rate of $\txnrate$ transactions/second, and their inter-generation times follow an exponential distribution. At each iteration of test runs, we not only vary the transaction generation rate $\txnrate$, but also measure the endorsing, ordering, validation, and ledger-commitment latency in order to analyze the impact of increasing $\txnrate$. Note that $\block$ and $\timeout$ refer to the block size and block-generation timeout, respectively.

\subsection{Latency Modeling}\label{subsec:Modeling}
For the latency modeling, we first obtain the histogram for each latency type defined in Section \ref{subsec:HLFLatencyTypes}, and then figure out its best-fit probability distribution. 
The discovered best-fit distributions are as follows.

\subsubsection{Endorsing Latency}
The best-fit distribution of the endorsing latency is the exponential distribution $\Te\sim\text{Exp$(\lambda)$}$, whose \ac{PDF} is defined as
\begin{equation}\label{eq:TePDF}
	f_{\Te}(t) ={\lambda}e^{-\lambda t},
\end{equation}
where $\lambda=\frac{1}{\expectation[\Te]}$ is the rate parameter for $\lambda>0$. Note that the \ac{CDF} of the endorsing latency is $F_{\Te}(t) =1-e^{-\lambda t}$.

\subsubsection{Ordering and Ledger-commitment Latencies}
The best-fit distributions of the ordering latency and ledger-commitment latency are the Gamma distribution, $\To\sim\text{Gamma$(\alphao,\betao)$}$ and $\Tt\sim\text{Gamma$(\alphat,\betat)$}$ respectively, whose \ac{PDF} is defined as
\begin{align}\label{eq:ToPDF}
	f_{\Ti}(t) =\frac{{\betai}^{\alphai}}{\Gamma(\alphai)}t^{{\alphai}-1}e^{-{\betai}t},
\end{align}
where $i$ is an indicator for the ordering latency $(i=\text{o})$ and for the ledger-commitment latency $(i=\text{t})$, $\Gamma(\alphai)=\int_{0}^{\infty} x^{\alphai-1} \,e^{-x} \,\text{d}x $ is the gamma function, and $\gamma(\alphai, \betai t)=\int_{0}^{\betai t} x^{\alphai-1} \,e^{-x} \,\text{d}x$ is the lower incomplete gamma function for $t\geq0$. Note that $\alphao=\betao \expectation[\To]$ and $\alphat=\betat \expectation[\Tt]$, where $\alphao$ and $\alphat$ are the shape parameters and $\betao$ and $\betat$ are the inverse scale parameters (i.e., the rate parameters). The \acp{CDF} of the ordering and ledger-commitment latencies are $F_{\Ti}(t) =\frac{1}{\Gamma(\alphai)}{\gamma({\alphai},{\betai}t)}$.

\begin{table}[!t]\scriptsize
	\caption{Estimated Gamma distribution parameters ($\alphat,\betat$),  
		the KS test results,
		the average latency of empirical results $\TVarep$, 
		and the average latency of the best-fit distribution $\TVarbf$ for the ledger-commitment latency.} 
	\begin{center}
		\renewcommand{\arraystretch}{1.5}
		%\resizebox{\textwidth}{!}{%
			\begin{tabular}{c|c|c|c|c|c|c}
				\specialrule{.1em}{.05em}{.05em}
				$\block$ & $\timeout$ & \textbf{$\txnrate$} 
				& {\makecell{\textbf{Estimated}\\ $(\alphat, \betat)$}} %$(\alpha_{txn}, \beta_{txn})$ 
				& {\makecell{\textbf{KS}\\ \textbf{Statistics}}} 
				& {\makecell{\textbf{Empirical}\\ \textbf{Avg.} $\TVarep$}} 
				& {\makecell{\textbf{Best-fit}\\ \textbf{Avg.} $\TVarbf$}}  \\
				\hhline{=======}
				
				\multirow{4}{*}{20} & \multirow{4}{*}{1}
				& 10 & (8.9573, 5.5858) & 0.0457 & 1.6066 & 1.6035 \\
				\cline{3-7}
				& & 11 & (9.5112, 6.0407) & 0.0466 & 1.5767 & 1.5745 \\
				\cline{3-7}
				& & 12 & (9.3159, 6.2078) & 0.0503 & 1.4969 & 1.5006 \\
				\cline{3-7}
				& & 13 & (10.0333, 6.2937) & 0.049 & 1.5891 & 1.5941 \\
				\specialrule{.1em}{.05em}{.05em}
				
				\multirow{4}{*}{25} & \multirow{4}{*}{1}
				& 17 & (9.7902, 6.0267) & 0.0504 & 1.6141 & 1.6244 \\
				\cline{3-7}
				& & 18 & (9.1368, 6.131) & 0.0595 & 1.4908 & 1.4902 \\
				\cline{3-7}
				& & 19 & (9.005, 5.9462) & 0.0564 & 1.5098 & 1.5144 \\
				\cline{3-7}
				& & 20 & (8.2069, 5.395) & 0.051 & 1.5232 & 1.5212 \\
				\specialrule{.1em}{.05em}{.05em}
				
				\multirow{4}{*}{10} & \multirow{4}{*}{2}
				& 8 & (7.181, 4.0151) & 0.0388 & 1.8289 & 1.7884 \\
				\cline{3-7}
				& & 9 & (6.8603, 4.6625) & 0.0437 & 1.4658 & 1.4713 \\
				\cline{3-7}
				& & 10 & (5.6829, 4.355) & 0.0457 & 1.3517 & 1.3049 \\
				\cline{3-7}
				& & 11 & (7.1636, 4.1801) & 0.0444 & 1.7513 & 1.7137 \\
				\specialrule{.1em}{.05em}{.05em}
				
				\multirow{4}{*}{20} & \multirow{4}{*}{2}
				& 15 & (7.6898, 4.6474) & 0.0358 & 1.6587 & 1.6566 \\
				\cline{3-7}
				& & 16 & (8.4486, 5.1794) & 0.0474 & 1.6406 & 1.6311 \\
				\cline{3-7}
				& & 17 & (8.0874, 5.2866) & 0.0419 & 1.5323 & 1.5297 \\
				\cline{3-7}
				& & 18 & (8.3269, 4.8638) & 0.0404 & 1.7437 & 1.712 \\
				\specialrule{.1em}{.05em}{.05em}
				
				\multirow{4}{*}{20} & \multirow{4}{*}{3}
				& 15 & (7.9739, 4.9723) & 0.0382 & 1.616 & 1.6036 \\
				\cline{3-7}
				& & 16 & (7.608, 4.9499) & 0.0394 & 1.5351 & 1.537 \\
				\cline{3-7}
				& & 17 & (7.0015, 4.9697) & 0.0403 & 1.3992 & 1.4088 \\
				\cline{3-7}
				& & 18 & (5.7078, 4.1482) & 0.0513 & 1.3902 & 1.3759 \\
				\specialrule{.1em}{.05em}{.05em}
				
				\multirow{4}{*}{30} & \multirow{4}{*}{3}
				& 23 & (8.0788, 4.6915) & 0.0369 & 1.727 & 1.722 \\
				\cline{3-7}
				& & 24 & (7.0546, 4.527) & 0.0395 & 1.5652 & 1.5583 \\
				\cline{3-7}
				& & 25 & (6.3698, 4.4257) & 0.0468 & 1.4345 & 1.4392 \\
				\cline{3-7}
				& & 26 & (7.3836, 4.8698) & 0.0435 & 1.5125 & 1.5162 \\
				\specialrule{.1em}{.05em}{.05em}
				
				9 & \multirow{4}{*}{2} & \multirow{4}{*}{10} & (8.9573, 5.5858) & 0.0457 & 1.6066 & 1.6035 \\
				\cline{1-1}\cline{4-7}
				10 & & & (9.5112, 6.0407) & 0.0466 & 1.5767 & 1.5745 \\
				\cline{1-1}\cline{4-7}
				12 & & & (9.3159, 6.2078) & 0.0503 & 1.4969 & 1.5006 \\
				\cline{1-1}\cline{4-7}
				15 & & & (10.0333, 6.2937) & 0.049 & 1.5891 & 1.5941 \\
				\specialrule{.1em}{.05em}{.05em}
				
				\multirow{4}{*}{10}
				& 1.75 & \multirow{4}{*}{10} & (9.7902, 6.0267) & 0.0504 & 1.6141 & 1.6244 \\
				\cline{2-2} \cline{4-7}
				& 2.00 & & (9.1368, 6.131) & 0.0595 & 1.4908 & 1.4902 \\
				\cline{2-2} \cline{4-7}
				& 2.25 & & (9.005, 5.9462) & 0.0564 & 1.5098 & 1.5144 \\
				\cline{2-2} \cline{4-7}
				& 2.50 & & (8.2069, 5.395) & 0.051 & 1.5232 & 1.5212 \\
				\specialrule{.1em}{.05em}{.05em}
			\end{tabular}
			\vspace{-5mm}
			%}
		\label{table:ModelingResults}
	\end{center}
\end{table}

\subsubsection{Validation Latency}
The best-fit distribution of the validation latency is the three-parameter Fr\'{e}chet distribution $\Tv\sim\text{Frechet$(\alpha,s,m)$}$, whose \ac{PDF} is defined as \cite{WuYouAbbAliKha:20}
\begin{equation}\label{eq:TvPDFCDF}
	f_{\Tv}(t)=\frac{\alpha}{s}\left(\frac{t-m}{s}\right)^{-1-\alpha} e^{-\left(\frac{t-m}{s}\right)^{-\alpha}},
\end{equation}
where $\alpha$, $s$, and $m$ are the shape, scale, and location parameters, respectively, for $0<\alpha<\infty$, $0<s<\infty$, and $-\infty<m<\infty$. Note that the \ac{CDF} of the validation latency is $F_{\Tv}(t)=e^{-\left(\frac{t-m}{s}\right)^{-\alpha}}$.

\begin{figure}[t!]
	\centering
	%\captionsetup{justification=centering}
	\begin{center}   
		{
			%\vspace{-3mm}
			%
			\psfrag{XXX}[Bc][bc][0.8] {$t$ [sec]}
			\psfrag{YYY}[Bc][bc][0.8] {Cumulative distribution function $F_{\Tt}(t)$}
			\psfrag{DDDDDDDDDDDDDDDDDDDl}[Bl][Bl][0.7] {Empirical results, $\txnrate=8$}
			\psfrag{data2}[Bl][Bl][0.7] {Best-fit distribution, $\txnrate=8$}
			\psfrag{data3}[Bl][Bl][0.7] {Empirical results, $\txnrate=9$}
			\psfrag{data4}[Bl][Bl][0.7] {Best-fit distribution, $\txnrate=9$}
			\psfrag{data5}[Bl][Bl][0.7] {Empirical results, $\txnrate=10$}
			\psfrag{data6}[Bl][Bl][0.7] {Best-fit distribution, $\txnrate=10$}
			\psfrag{data7}[Bl][Bl][0.7] {Empirical results, $\txnrate=11$}
			\psfrag{data8}[Bl][Bl][0.7] {Best-fit distribution, $\txnrate=11$}
			\includegraphics[width=1.00\columnwidth]{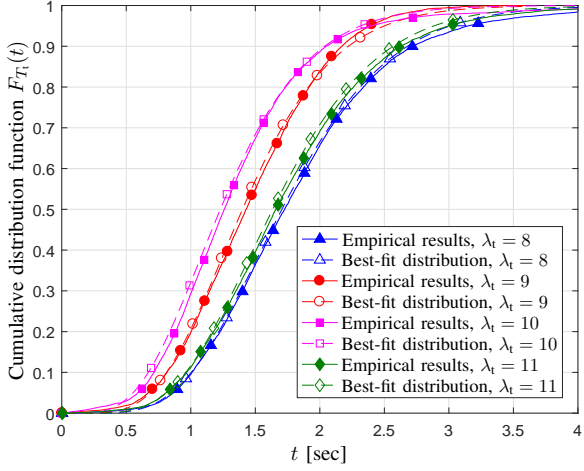}
			%			\vspace{-10mm}
		}
	\end{center}
	\caption{
		Empirical CDFs and the best-fit Gamma distribution CDFs of 
		the ledger-commitment latency $\Tt$
		for different values of the transaction generation rates $\txnrate$, 
		where $\block=10$ and $\timeout=2$ [sec].
	}
	\vspace{-3mm}
	\label{fig:modelingcdf}
\end{figure}

Once the best-fit distributions are determined, we estimate the distribution parameters, such as $\lambda$ for the endorsing, $(\alpha, s, m)$ for the validation, and $(\alphao, \betao)$ and $(\alphat, \betat)$ for the ordering and ledger-commitment latencies, using the curve fitting tool such as the nonlinear regression method in Matlab.\footnote{For better fitting results, we exclude some test runs with remarkably long mean latency, similarly to conventional fitting work such as \cite{SukMarChaTriRin:17}.} The histogram and discovered best-fit distribution with estimated parameter for each latency type are provided in Fig. \ref{fig:lambda10}. Note that $\Deltat$ is the width of a bin. From Fig. \ref{fig:lambda10}, we can see that the discovered best-fit distributions match well with the histograms, especially with that of the ledger-commitment latency.

For the ledge-commitment latency $\Tt$, the estimated Gamma distribution parameter pairs of the ledger-commitment latency for various \ac{HLF} environments are also provided in Table \ref{table:ModelingResults}. The \acp{CDF} of some empirically measured values and their best-fit Gamma distributions are then demonstrated in Fig.~\ref{fig:modelingcdf}. Note that the value of $\Deltat$ does not change the shapes of histograms, so we can still use the estimated Gamma distribution parameters in Table \ref{table:ModelingResults}, even for different values of $\Deltat$. In Table \ref{table:ModelingResults}, $\TVarAvg$ refers to the average ledger-commitment latency for a given \ac{HLF} environment. The one, obtained empirically, is denoted as $\TVarep$, and the one from the best-fit distribution is denoted as $\TVarbf$.

From Fig. \ref{fig:modelingcdf}, we can see that the \acp{CDF} of the best-fit Gamma distribution matches well with the empirically obtained ones. From Table \ref{table:ModelingResults}, we can see that $\TVarAvg$ decreases as $\txnrate$ increases since it makes new blocks generated faster in the ordering service. However, $\TVarAvg$ increases when $\txnrate$ becomes higher. This is because there is a large load at the validation phase as the block-generation rate becomes higher. Hence, some blocks start to wait for the validation process, which increases $\TVarAvg$.

\subsection{Latency Modeling Validation}\label{subsec:ModelingValidation}
In order to validate and improve the reliability of our latency model, we include \ac{KS} test results in Table \ref{table:ModelingResults}. The \ac{KS} test is a statistical test to evaluate goodness-of-fit based on the maximum difference between the empirical \ac{CDF} of the data set and the hypothetical \ac{CDF} (i.e., the \ac{CDF} of the given distribution) \cite{PreTeu:88}.

The estimated parameters and KS statistics are averaged over ten test runs at 0.01 significance level. The critical value is computed as 0.0513. In other words, the latency model can be considered as reliable if the KS statistics value of a data set is less than 0.0513. As an example, Fig. \ref{fig:modelingcdf} show the ledger-commitment latency, when $\block=10$ and $\timeout=2$ [sec] for $\txnrate=8, 9, 10$ and $11$. In Table \ref{table:ModelingResults}, all cases in those figures have smaller \ac{KS} statistics values (i.e., 0.0388, 0.0437, 0.0457, and 0.0444, respectively) than the critical value. Thus, the test is passed for all cases. However, in Table \ref{table:ModelingResults}, when $\block=25$ and $\timeout=1$ for $\txnrate=18$ and $19$, their KS statistics values are slightly larger than the critical value, but even in those cases, we can see that they still have the similar mean latency $\TVarbf$ to each of their $\TVarep$.

\subsection{Feasibility Conditions for Latency Modeling}\label{subsec:ModelingFeasibleConditions}
During the modeling process, it is found that there exist some cases in which histograms of $\TVarAvg$ do not fit well into a known probability distribution. 
In other words, the probability distribution fitting method is not always applicable.
We have discovered two environments that make the probability distribution fitting method infeasible as follows.

\subsubsection{Timeout-dominant Block Generation Environment}
When the transaction generation rate $\txnrate$ is small, while the block-generation timeout value $\timeout$ is relatively large, few of transactions (or even one transaction) may stay in the waiting room for the ordering service in the ordering phase most of their time, and become a block once the timeout expires.
In this case, a large portion of transactions can have the ordering latency equal to $\timeout$. As an example of such case, the histograms of the ordering latency and the ledger-commitment latency are shown in Figs. \ref{fig:ol_t1b10r3} and \ref{fig:lc_t1b10r3}, respectively, where $\block=10$, $\timeout=1$ [sec], and $\txnrate=3$. As shown in Fig. \ref{fig:ol_t1b10r3}, there is a peak when the ordering latency is 1.2 seconds, 
and in this case, the percentage of blocks, which are exported from the ordering service by the block-generation timeout expiration, is 99.8\%. Due to this, as shown in Fig. \ref{fig:lc_t1b10r3}, the ledger-commitment latency has the distribution, which is difficult to fit using any known probability distributions.

\subsubsection{Block Size-dominant Block Generation Environment}
\begin{figure*}[t!]
	\centering
	\begin{subfigure}[b]{0.325\textwidth}
		\centering
		\psfrag{XXX}[Bc][bc][0.8] {$t$ [sec]}
		\psfrag{YYY}[Bc][bc][0.8] {Probability/$\Deltat$}
		\includegraphics[width=\textwidth]{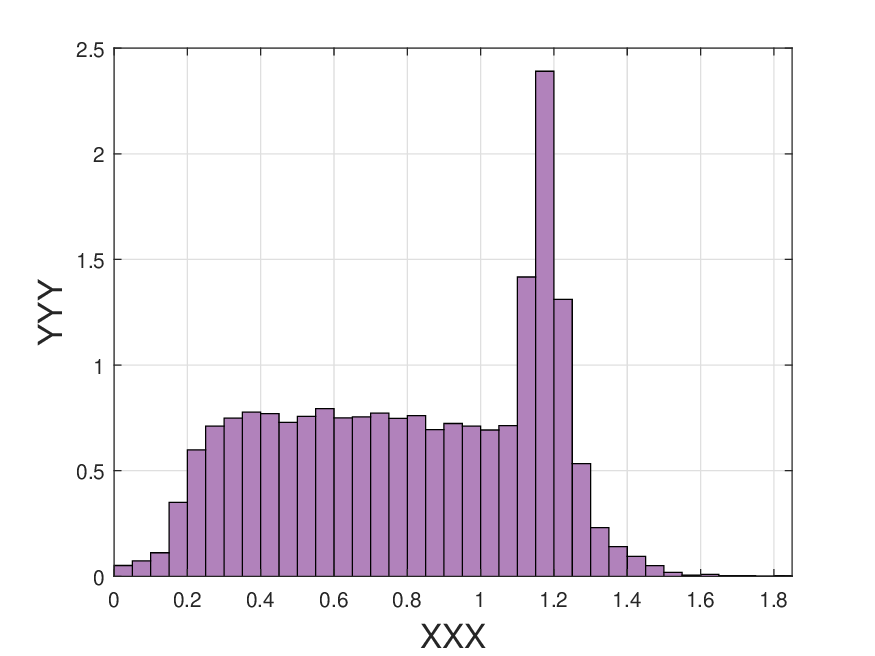}
		\caption{Ordering latency}
		\label{fig:ol_t1b10r3}
	\end{subfigure}
	\hfill
	\begin{subfigure}[b]{0.325\textwidth}
		\centering
		\psfrag{XXX}[Bc][bc][0.8] {$t$ [sec]}
		\psfrag{YYY}[Bc][bc][0.8] {Probability/$\Deltat$}
		\includegraphics[width=\textwidth]{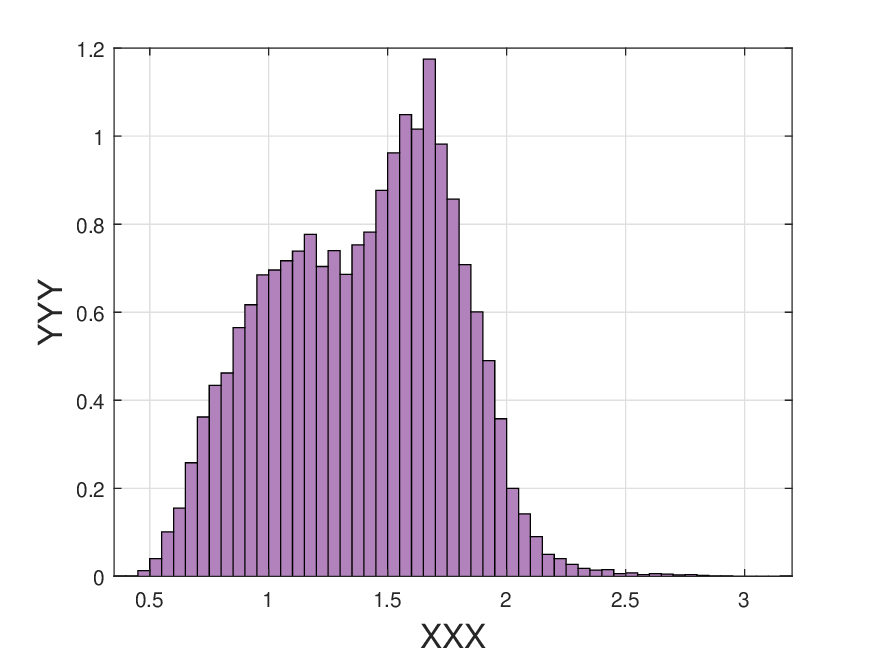}
		\caption{Ledger-commitment latency}
		\label{fig:lc_t1b10r3}
	\end{subfigure}
	\hfill
	\begin{subfigure}[b]{0.325\textwidth}
		\centering
		\psfrag{XXX}[Bc][bc][0.8] {$t$ [sec]}
		\psfrag{YYY}[Bc][bc][0.8] {Probability/$\Deltat$}
		\includegraphics[width=\textwidth]{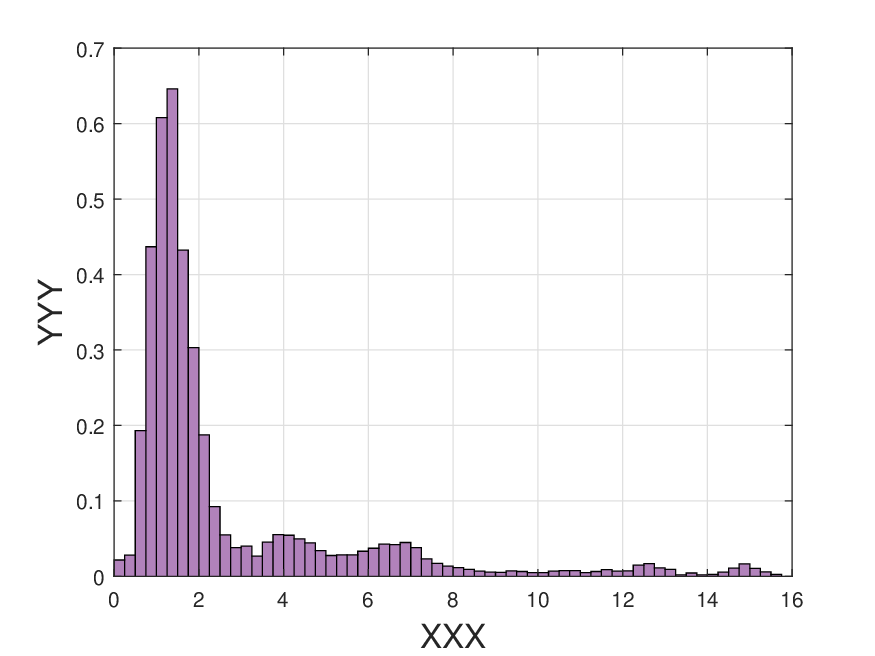}
		\caption{Ledger-commitment latency}
		\label{fig:lc_t075b15r10}
	\end{subfigure}
	
	\caption{Histograms of the ordering latency and the ledger-commitment latency $\Tt$ for infeasible latency modeling cases.}
	\vspace{-3mm}
	\label{fig:modelingCondition}
\end{figure*}

When the block size $\block$ is small, while the transaction generation rate $\txnrate$ is large, 
most blocks are filled with transactions up to $\block$ quickly, 
so blocks are generated fast with the full size from the ordering service.
In this case, even though the ordering latency is small, there can be a large load at the validation phase due to the fast block generation, which makes blocks wait for validation.
Hence, there can be many transactions with long waiting time to enter the validation phase. 
As an example of such case, the histogram of the ledger-commitment latency is provided in Fig.~\ref{fig:lc_t075b15r10},
where $\block=15$, $\timeout=0.75$ [sec], and $\txnrate=10$.
As shown in this figure, there are long-tail outliers more frequently in the distribution (e.g., the cases with the ledger-commitment latency longer than 4 seconds), which makes it difficult to fit to a known probability distribution.

Note that the two environments, discussed above, are the cases where $\block$ or $\timeout$ is too small or large, so a large portion of transactions experience long latency in the ordering or validation phase. For delay-sensitive \ac{HLF}-enabled \ac{IoT}, such \ac{HLF} parameter setups will be generally avoided.

\section{Facing to Blockchain Latency: Influential Parameter Analysis on Latency}\label{sec:ParameterAnalysis}
For \ac{HLF}-enabled \ac{IoT} handling time-sensitive jobs and requiring faster transaction commitment, it is important to design \ac{HLF} in the way to minimize the ledger-commitment latency. We can see in Table \ref{table:ModelingResults}, $\block$, $\timeout$, and $\txnrate$ affect the ledger-commitment latency. In this section, therefore, we explore their impacts on the ledger-commitment latency. Moreover, we discuss how to optimally determine the three parameters to minimize the ledger-commitment latency.

\subsection{Transaction Generation Rate Control}\label{subsec:TxnRateImpact}
Table \ref{table:ModelingResults} demonstrates the impact of the transaction generation rate $\txnrate$ for different $\block$ and $\timeout$ on the average ledger-commitment latency $\TVarAvg$ for both empirical and best-fit distribution results. From Table \ref{table:ModelingResults}, we can observe two conflicting effects of the transaction generation rate on the ledger-commitment latency. Higher $\txnrate$ makes transactions exported from the ordering service faster, leading to an increase in the block-generation rate at the ordering service. Due to this, the lowest $\TVarAvg$ can be achieved by increasing $\txnrate$. For instance, when $\block$ is $20$ and $\timeout$ is $1$ [sec] in Table \ref{table:ModelingResults}, $\TVarep$ decreases from 1.6066 to 1.4969 as $\txnrate$ increases from 10 to 12 transactions/second.

However, $\TVarAvg$ does not continuously decrease, and starts to increase after some point of $\txnrate$. As seen in Table \ref{table:ModelingResults}, when $\block$ is $20$ and $\timeout$ is $1$ [sec], $\TVarep$ increases from 1.4969 to 1.5891 as $\txnrate$ increases from $12$ to $13$ transactions/second. This is because the block validation rate does not increase despite faster block-generation. New blocks need to wait for the validation. From these observations, we can see that the optimal $\txnrate$, which minimizes $\TVarAvg$, exists for a given \ac{HLF} setup.

\subsection{Block Size Optimization}\label{subsec:BlocksizeImpact}
Figure \ref{fig:parameterImpacts} illustrates the impact of the block size $\block$ on $\TVarAvg$ for empirical results and the best-fit distributions (red lines).
Note that the probability distribution fitting results with \ac{KS} statistics values are provided in Table \ref{table:ModelingResults} (see the values for $\block=9, 10, 12,$ and $15$ and $\timeout=2$ [sec]). From Fig. \ref{fig:parameterImpacts}, we can see that the average ledger-commitment latency of the best-fit distributions are well-matched with those of the empirical results.

From Fig. \ref{fig:parameterImpacts}, we can also see that when the block size $\block$ is small (e.g., $\block<8$), the average ledger-commitment latency $\TVarAvg$ is extremely high. This is because the peers receiving blocks cannot promptly start to validate them since blocks are generated too fast due to the small block size.
Therefore, new blocks gradually stack up in each peer's queue for the validation, which leads to long validation delay. 
As $\block$ increases, $\TVarAvg$ decreases because the block-generation rate decreases and the waiting time at the validation phase becomes lower as well. 
However, when $\block$ increases beyond certain values (e.g., $\block>10$), $\TVarAvg$ increases again. 
In this case, the validation latency is no longer long, but transactions need to wait longer for others to be included into a block in the ordering phase due to the large size of $\block$. As $\block$ keeps increasing, $\TVarAvg$ could eventually saturate as most of the new blocks are generated by the timeout expiration before they are generated by fully collected $\block$ transactions.

\subsection{Block-generation Timeout Optimization}\label{subsec:TimeoutImpact}
\begin{figure}[t!]
	\centering
	%\captionsetup{justification=centering}
	\begin{center}   
		{
			\psfrag{XXX1}[Bc][bc][0.8] {\color[RGB]{255, 0, 0} Block size $\block$ [number]}
			\psfrag{XXX2}[Bc][bc][0.8] {\color[RGB]{0, 0, 255} Block-generation timeout $\timeout$ [sec]}
			\psfrag{YYY}[Bc][bc][0.8] {Average ledger-commitment latency $\TVarAvg$ [sec]}
			\psfrag{a}[Bl][Bl][0.7] {Empirical results ($\block$ varies, $\timeout$ = 2)}
			\psfrag{b}[Bl][Bl][0.7] {Best-fit distribution ($\block$ varies, $\timeout$ = 2)}
			\psfrag{c}[Bl][Bl][0.7] {Empirical results ($\block$ = 10, $\timeout$ varies)}
			\psfrag{d}[Bl][Bl][0.7] {Best-fit distribution ($\block$ = 10, $\timeout$ varies)}
			\includegraphics[width=1.00\columnwidth]{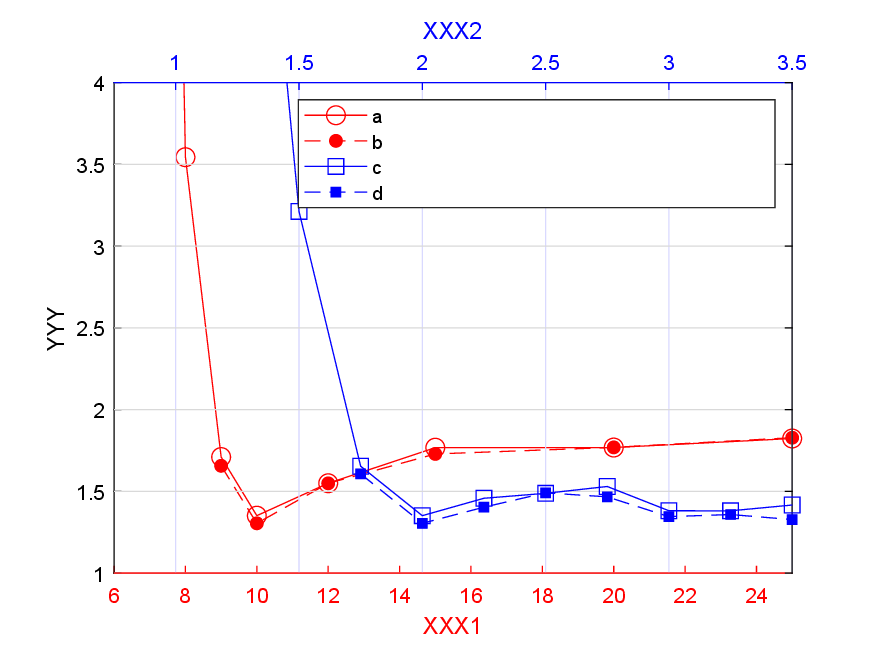}
			%			\vspace{-10mm}
		}
	\end{center}
	\caption{
		Effects of the block size $\block$ and block-generation timeout $\timeout$ on the average ledger-commitment latency $\TVarAvg$ for $\txnrate=10$, where $\block$ varies and $\timeout=2$, and where $\block=10$ and $\timeout$ varies, respectively.
	}
	\vspace{-3mm}
	\label{fig:parameterImpacts}
\end{figure}

Figure \ref{fig:parameterImpacts} demonstrates the impact of the block-generation timeout $\timeout$ on $\TVarAvg$ for empirical results and the best-fit distributions (blue lines). Note that the probability distribution fitting results with \ac{KS} statistics values are provided in Table \ref{table:ModelingResults} (see the values for $\block=10$ and $\timeout=1.75, 2.00, 2.25,$ and $2.50$ [sec]).
From Fig \ref{fig:parameterImpacts}, we can see that the average ledger-commitment latency of the best-fit distributions are well-matched with those of the empirical results.

From Fig. \ref{fig:parameterImpacts}, we can also see that when the block-generation timeout $\timeout$ is short (e.g., $\timeout<1.5$ [sec]), the average ledger-commitment latency $\TVarAvg$ is extremely high for the same reason as explained in Section \ref{subsec:BlocksizeImpact} due to the short block-generation timeout. As $\timeout$ increases, $\TVarAvg$ decreases and the waiting time at the validation phase also becomes lower.
However, when $\timeout$ increases beyond certain values (e.g., $\timeout<2$ [sec]), $\TVarAvg$ increases again. In this case, the validation latency is no longer long, but transactions need to wait longer for others to be included into a block in the ordering phase due to the long $\timeout$. As $\timeout$ keeps increasing, $\TVarAvg$ could eventually saturate as most of the new blocks are generated because $\block$ transactions are collected before the timeout expires.

Besides, we found that $\TVarAvg$ can be more sensitively changed according to $\txnrate$ for longer $\timeout$. Specifically, in Table \ref{table:ModelingResults}, for $\block=20$, when $\timeout$ is $2$ [sec], $\TVarep$ is 1.6587 and 1.7437, respectively, for $\txnrate=15$ and $18$. However, when $\timeout$ becomes $3$ [sec], $\TVarep$ is 1.616 and 1.3902, respectively, for $\txnrate=15$ and $18$. The difference of $\TVarep$ is larger for $\timeout=3$ [sec] (i.e., 0.2258) than that for $\timeout=2$ [sec] (i.e., 0.085).

\section{Conclusions}\label{sec:Conclusions}
This article provides the latency model for \ac{HLF}-enabled \ac{IoT}, where time-sensitive data is generally managed. From the histograms of the ledger-commitment latency obtained from our real implementation of \ac{HLF}, we figure out that the ledger-commitment latency can be modeled as the Gamma distribution in various \ac{HLF} setups. We also conduct the \ac{KS} test and verify that our modeling is reliable. Moreover, we explore the impacts of three important \ac{HLF} parameters (i.e., transaction generation rate, block size, and block-generation timeout) on the ledger-commitment latency. Specifically, when they are set to be small, the ledger-commitment latency is fairly high due to the fast block generation that causes long validation latency. Hence, the ledger-commitment latency decreases as those values increase, but eventually it increases again since the latency in the ordering phase increases. From those observations, it is shown that a proper setup of \ac{HLF} parameters are important to lower the ledger-commitment latency. This article can serve as a useful framework not only to predict the mean and distribution of the ledger-commitment latency, but also to optimally design \ac{HLF}-enabled \ac{IoT} with low latency.

\bibliographystyle{bib/IEEEtran}

\end{document}